\documentclass[pra,superscriptaddress, nofootinbib, showpacs ]{revtex4}
\usepackage{graphicx}       % for eps figures
\usepackage{bm}         % for bold math symbols
\usepackage{amsmath}        % for \text and such
\usepackage{latexsym}
\begin{document}

\title{Entropic uncertainties for joint quantum measurements}
\author{Thomas Brougham$^{1,2}$, Erika Andersson$^3$ and Stephen M Barnett$^2$\\
$^1$Department of Physics, FJFI, \v{C}VUT, Praha, Czech Republic\\
$^2$SUPA, Department of Physics, University of Strathclyde, Glasgow G4 ONG, UK\\
$^3$SUPA, Department of Physics, School of EPS, Heriot-Watt University, Edinburgh}
%\affiliation{SUPA, Department of Physics, University of Strathclyde, Glasgow G4 0NG, UK}
%\affiliation{SUPA, Department of Physics, University of Strathclyde, Glasgow G4 ONG, UK}
\date{\today}

\begin{abstract}
We investigate the uncertainty associated with a joint quantum measurement of two spin components of a spin-1/2 particle and quantify this in terms of entropy.  We consider two entropic quantities: the joint entropy and the sum of the marginal entropies, and obtain lower
bounds for each of these quantities.  For the case of joint measurements where we measure each spin observable equally well, these lower bounds are tight. 
\end{abstract}
\maketitle

\section{Introduction}
One of the most striking features of quantum mechanics is that there are fundamental limits to how well we can measure quantities.  For instance, an atom can be prepared so that it has either a well localized position or has a low spread in momentum. We cannot, however, prepare it so that it has both a well localized position and a low spread in momentum.  In terms of measurement this means that a system cannot be prepared so that a measurement of either position or momentum would both have low uncertainty.  This is a general feature of quantum measurements and applies to any non-commuting observables that do not share a common eigenvector.  This fact is embodied mathematically by the uncertainty principle, the most common form of which is \cite{rob} %While it is possible, in principle, to prepare a system so that there is no uncertainty in the outcome of a measurement of a single observable, it is not possible to prepare the system so that all measurements have no uncertainty in their outcome.  In particular if we have two non-commuting observables and we will  going to measure on of them on the system, then we can not generally prepare the system so that there is no uncertainty in the outcome. This last fact is described mathematically by the uncertainty principle, which all This can be illustrated by the following example.  If The intrinsic limitations on measuring the properties of two non-commuting observables are embodied in the uncertainty principle, the most common form of which is \cite{rob}
\begin{equation}
\Delta\hat A\Delta\hat  B\ge1/2|\langle[\hat A,\hat B]\rangle|.
\end{equation}
The lower bound on this uncertainty product involves an expectation value and is thus state dependent.  This can make it difficult to bound the uncertainty; in particular it is possible
for the lower bound to vanish.  For example, the observables $\hat \sigma_x$ and $\hat \sigma_y$ do not commute, however the lower bound on the uncertainty will be zero if our system is prepared in the maximally mixed state, i.e. $\hat \rho=\hat 1/2$.  These difficulties have led some authors to consider entropic uncertainty principles \cite{hirsch,bialynicki,deutsch, maassen}.  These have the advantage of
having non-trivial state independent lower bounds.  Both the standard uncertainty principle \cite{rob} and the entropic uncertainty principles \cite{bialynicki,deutsch, maassen} are derived under the assumption that the observables are measured on separate but
identically prepared systems.  They do not apply to the case where both quantities are measured at the same time on a single system.  %It is sometimes thought that joint measurements of two non-commuting observables is impossible, this is not the case.      

In this paper we will investigate entropic uncertainty relations for joint measurements of two spin components of a spin-1/2 particle.  A joint measurement is a single
observation that gives information about two different observables.  If two observables do not commute, then we cannot jointly measure them using projective measurements alone.  Instead we must use generalized measurements, which can add extra uncertainty to the measurements \cite{massar}.  It seems that the price one pays for jointly measuring two observables is that one necessarily increases the uncertainty in the measurement process.  This would appear to fit with the general feature of quantum mechanics to limit the amount of information that can be obtained from a system.  This leads naturally to the question: how much information can be extracted from a single system concerning two observables?  To address this we must first determine how well we can measure the observables. 

%The process of jointly measuring two non-commuting observables $\hat A$ and $\hat B$ can be understood in the following manner.  In any real measurement of $\hat A$ we will have experimental uncertainty as well as the intrinsic quantum uncertainty.  This means that the statistics of a measurement of $\hat A$ will not be the same as those found by projecting onto the eigenvectors of $\hat A$.  Instead, we can describe the statistics by using a generalised quantum measurement \cite{nch, peres}.  Let $A'$ and $B'$ be two generalized measurements that approximate $\hat A$ and $\hat B$ respectively.  When the uncertainty in $A'$ and $B'$ is sufficiently large it is possible to make a single measurement that simultaneously measures $A'$ and $B'$ \cite{AK, AG}.  This single measurement will provide information about both $\hat A$ and $\hat B$, hence it can be interpreted as a joint measurement of these two observables.  %The price one pays, however, for jointly measuring the two observables is that one necessarily increases the uncertainty in the measurement process.

An important question to ask in relation to joint measurements is how much extra uncertainty we have compared to when we measure each observable separately on identically prepared systems.  An uncertainty principle for joint measurements would help in answering this question as it would give bounds on the uncertainty inherent in the process of jointly measuring two observables.  The study of how well we can jointly measure two observables is also of interest with regards to assessing the security of quantum cryptography protocols \cite{crypto}.

%An uncertainty principle for joint measurements will thus give information both about the uncertainty that arises from complementarity, as well as giving an indication of how much uncertainty must be added when two observables are jointly measured.  
Variance based uncertainty relations, which use the product of the variance as a measure of uncertainty, have been obtained
for joint measurements \cite{AK,AG,erikas}.  We shall use, instead, the entropy of the measurement outcomes as a means of quantifying the uncertainty in a joint measurement.  One reason for this is that using entropy will lead to state independent lower bounds.  In addition to this entropy can be interpreted as a measure of information.  Two different entropic quantities will be investigated,
the sum of the marginal entropies and the joint entropy.  Lower bounds will be obtained for both quantities.

The paper is organised in the following manner.  A brief review of joint measurements will be given in section \ref{jmeasurements}.  In section \ref{eups} we will discuss entropic uncertainty principles.  A lower bound will be found in section \ref{jents} for the joint entropy of a joint measurement of two spin components of a spin 1/2 particle.  In section \ref{margents} we will find lower bounds on the sum of the marginal entropies for the joint measurement of two spin componets of a spin 1/2 particle.  Finally we will discuss the results in section \ref{conc}.

\section{Joint measurements}
\label{jmeasurements}
A joint measurement of two observables is a simultaneous measurement of both observables upon the same quantum system.  When the observables of interest commute
then a joint measurement can be accomplished with standard von Neumann or projective quantum measurements.  If the observables do not commute, however, then we must
describe our measurements in terms of the probability operator measure (POM) formalism \cite{walker,stig}.  A detailed description of this generalized description of measurements
can be found in \cite{nch, peres}.  An important point to note is that any measurement can be described using a POM.  %In particular the standard description of measurement corresponds to the case when our probability operators are just the projectors onto the eigenvalues of the observable.  
Conversely, any POM can be realized in terms of a projective measurement on a larger system\footnote{To see how this might arise one needs only to consider a realistic description of the measurement process.  An experiment consists not only of the system of interest, but also of a measurement apparatus that interacts with the system.  The apparatus is a physical object and thus can be described quantum mechanically.  The net effect of this is that we make a projective measurement on the extended system, which can be represented by a POM acting on the initial system.} \cite{aperes}. 

A condition that is frequently used in connection with joint measurements is the joint unbiasedness condition \cite{AG, busch}.  This condition requires that the expectation values for the jointly measured
observables are proportional to the expectation values of the observables measured separately.  We thus require that $\langle\hat A_{J}\rangle=\alpha\langle\hat A\rangle$ and $\langle\hat B_{J}\rangle=\beta\langle\hat
B\rangle$, where $\langle\hat A_{J}\rangle$ and $\langle\hat B_{J}\rangle$ are the jointly measured expectation values and $\alpha$ and $\beta$ are constants of
proportionality, the values of which are independent of the state to be measured.  For examples of the joint unbiasedness condition applied to spin-1/2 systems see \cite{busch, busch2, demuynck, steves, erikas}.  It is possible to relax this
condition, leading to a more general description of joint measurements \cite{hall, teiko, coexist1, coexist2, jmqubit}.  In the remainder of this work we shall however consider only joint measurements that satisfy the unbiasedness condition.

We consider joint measurements of two spin components for a spin-1/2 particle. Let the spin observables that we seek to measure be $\hat A={\bf a\cdot}\boldsymbol{\hat\sigma}$ and $\hat B={\bf
b\cdot}\boldsymbol{\hat\sigma}$.  The eigenvalues of these observables are $\pm 1$.  We will choose to assign the numerical values of $\pm 1$ to the results spin up and down also for a joint measurement of $\hat A$
and $\hat B$.  This means that $|\alpha|$ and $|\beta|$ will vary between one and zero.  It can be shown that the values of $\alpha$ and $\beta$ will be restricted by the inequality \cite{erikas, busch}
\begin{equation}
\label{ineq}
 |\alpha{\bf a}+\beta{\bf b}|+|\alpha{\bf a}-\beta{\bf b}|\le 2.
\end{equation}
If a joint measurement scheme allows us to saturate inequality (\ref{ineq}), then, for given directions {\bf a} and {\bf b}, this measurement gives 
the largest possible value of  $|\alpha|$ for a given $\beta$ and vice versa.  Any joint measurement of two components of spin  for which inequality  
(\ref{ineq}) is saturated is in this sense an optimal joint measurement.

\begin{figure}
\center{\includegraphics[width=9cm,height=!]
{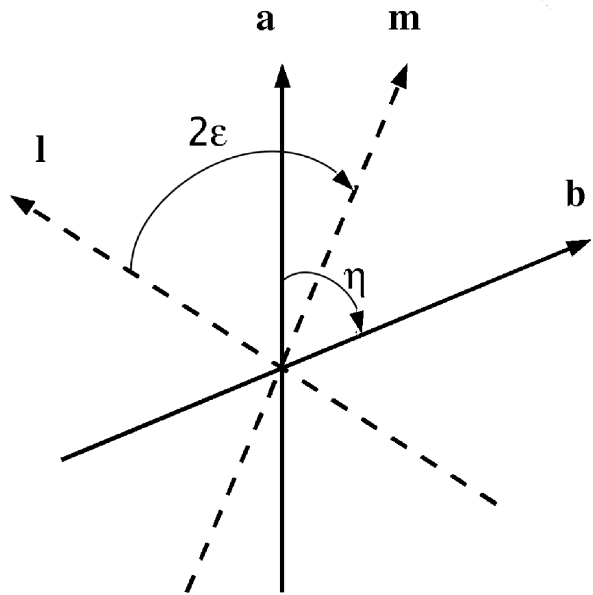}}
\caption{A diagram showing the ${\bf m}$ and ${\bf l}$ axes in relation to the ${\bf a}$ and ${\bf b}$ axes.}
\label{fig1}
\end{figure}

A simple scheme for performing an optimal joint measurement of two spin components of a spin-1/2 particle is given in \cite{busch2, erikas}. In this scheme we introduce two new spin axes ${\bf m}$ and
${\bf l}$, which we measure along.  The orientation of the ${\bf m}$ and ${\bf l}$ axes relative to ${\bf a}$ and ${\bf b}$ is shown in figure \ref{fig1}.  We will either
measure along ${\bf m}$, with a probability $p$, or along ${\bf l}$ with a probability $1-p$.  If we measure along ${\bf m}$, then the result `spin up' is interpreted as both
${\bf a}$ and ${\bf b}$ being spin up, and the result `spin down' along ${\bf m}$ is taken as ${\bf a}$ and ${\bf b}$ both spin down.  If instead we measured along ${\bf
l}$ then the result `spin up' is taken as ${\bf a}$ being spin up and ${\bf b}$ being spin down.  If we get the result `spin down' along ${\bf l}$ then we take ${\bf a}$ to be
spin down and ${\bf b}$ to be spin up.  This measurement scheme thus provides simultaneous results for the two observables $\hat A$ and $\hat B$.  

It can shown that this measurement scheme can be used to implement an optimal joint measurement when we choose ${\bf m}$, ${\bf l}$ and $p$ as  \cite{busch2, erikas}
\begin{eqnarray}
\label{m}
{\bf m}&=&\frac{1}{2p}\left(\alpha{\bf a}+\beta{\bf b}\right),\\
\label{l}
{\bf l}&=&\frac{1}{2(1-p)}\left(\alpha{\bf a}-\beta{\bf b}\right),\\
\label{p}
p&=&\frac{1}{2}|\alpha{\bf a}+\beta{\bf b}|.
\end{eqnarray}
For more details on how to implement an optimal joint measurements of spin see \cite{busch2, demuynck, erikas}.  An alternative method of performing an optimal joint measurement of two spin components
is to use quantum cloning \cite{cloning}.

We have taken a joint measurement of two observables to be a single measurement that gives us information about both observables.  This approach to joint measurements has been discussed by Uffink who has proposed an alternative definition of a joint measurement \cite{uffink}.  
%The description of joint measurements that we have outlined has been discussed by Uffink \cite{uffink}.  
In the current formalism, if we want to jointly measure $\hat A$ and $\hat B$, then we seek a POM $\hat\Pi^{ab}_{ij}$ such that $\hat\Pi^a_i=\sum_j{\hat\Pi^{ab}_{ij}}$ and $\hat\Pi^b_j=\sum_i{\hat\Pi^{ab}_{ij}}$, where $\hat\Pi^a_j$ and $\hat\Pi^b_j$ are unsharp measurements of the observables $\hat A$ and $\hat B$ respectively. It has been argued by Uffink that if $\hat\Pi^{ab}_{ij}$ is not a function of either $\hat A$ or $\hat B$, then it should not be interpreted as a true joint measurement of these observables. The justification, however, for how the term ``joint measurement of $\hat A$ and $\hat B$" is commonly used, is that the unsharp marginal observables approximate the two sharp observables. Performing the joint measurement thus gives us information about both observables. For instance, one can use joint measurements to estimate the expectation values of two sharp observables \cite{estimation}.  One could still argue, as Uffink does, that we are not actually jointly measuring $\hat A$ and $\hat B$. If we, however, apply this criterion consistently, then it leads to problems not only with joint measurements but with standard measurements.  To see this we note that in any real measurement there will be experimental uncertainty in addition to the intrinsic quantum uncertainty.  A measurement of an observable is thus not really measuring the observable. Instead we measure some POM that approximates the desired observable.  It is still common in physics to call this a measurement of the observable.

\section{The entropic uncertainty in a joint measurement}
\label{eups}
Standard entropic uncertainty principles use the sum of the marginal entropies to quantify the uncertainty in a measurement.  For example if we are measuring the observables $\hat A$ and $\hat B$, then the
total uncertainty would be 
\begin{equation}
H(A)+H(B)=-\sum_i{p^a_i\log(p^a_i)}-\sum_j{p^b_j\log(p^b_j)},
\end{equation}
where $p^a_i$ is the probability of obtaining the $i^{th}$ outcome for a measurement of $\hat A$ and $p^b_j$
is the probability of obtaining the $j^{th}$ outcome for a measurement of $\hat B$.  For the case when the observables have $d$ non-degenerate outcomes,  a strong lower bound on the sum of the entropies was found by Maassen and
Uffink \cite{maassen}.  Their result states that for two non-degenerate observables $\hat A$ and $\hat B$, with eigenvectors $\{|a_i\rangle\}$ and $\{|b_j\rangle\}$ respectively, we have that 
\begin{equation}
\label{mu}
H(A)+H(B)\ge \max_{i,j}\left(-2\log|\langle a_i|b_j\rangle|\right).
\end{equation}
In certain situations one can find an entropic uncertainty principle that is stronger than that given in equation (\ref{mu}).  For example, a stronger uncertainty principle exist for measurements of two spin components of a spin 1/2 particle \cite{GMR, sanchezruiz}.  Other extensions to the work of Maassen and Uffink have been found \cite{sruiz, winter}.  It is important to note that the entropic uncertainty principle of Maassen and Uffink does not apply to joint measurements of $\hat A$ and $\hat B$.  Rather, it was derived under the
assumption that the observables would be measured on separate but identically prepared systems.  

If we wish to quantify the uncertainty in a joint measurement of two observables $\hat A$ and $\hat B$, then we could use the sum of the marginal entropies $H(A)+H(B)$, or the joint entropy $H(A,B)$.  It is common to use the sum of the marginal entropies as a measure of uncertainty \cite{deutsch,maassen}.  To facilitate a comparison between our results and the existing literature we shall investigate the sum of the marginal entropies.  The joint entropy, however, is also of interest for two reasons.  Firstly it deals directly with the joint probability, $P(A_J,B_J)$.  The importance of this is that the existence
of the joint probability distribution is a defining characteristic of joint measurements.  The second reason why the joint entropy should be considered is that any lower bound on the joint
entropy will also act as a lower bound on the sum of the marginal entropies.  This can be seen by recalling the fact that entropy is subadditive, which means that $H(A)+H(B)\ge H(A,B)$ \cite{wehrl, EIT}.  It is important to note that nontrivial state independent lower bounds can be obtained for both $H(A,B)$ and $H(A)+H(B)$.  These lower bounds will thus give us information about the minimum uncertainty that is inherent in the quantum mechanical measurement.  We shall now look at obtaining lower bounds on both the sum of the marginal entropies and the joint entropy for a joint measurement of two observables on a qubit.  A qubit is a spin-1/2 particle and thus we will consider a joint measurement of spin along two different directions.  Let our spin observables be $\hat A={\bf
a}\cdot\boldsymbol{ \hat \sigma}$ and $\hat B={\bf b}\cdot\boldsymbol{ \hat \sigma}$, where ${\bf a}$ and ${\bf b}$ are both unit vectors.  We consider the entropy of an optimal joint measurement
of spin, i.e. a measurement that saturates the inequality (\ref{ineq}).  The reason for this is that optimal joint measurements provide the largest possible values for the sharpnesses $\alpha$ and $\beta$ and this thus leads to
probability distributions with lower entropy.  For example if $\alpha$ is the sharpness of a measurement of $\hat A$, which is measured on a state $\hat\rho=1/2(1+{\bf c}\cdot\boldsymbol{\hat\sigma})$, then the
probability of obtaining the results spin up for a measurement of $\hat A$ is $P^a_+=1/2(1+\alpha{\bf a\cdot c})$.  It can thus be seen that as $\alpha$ becomes closer to 1, then $H(A)$ will decrease. 
This shows that taking the sharpnesses of the joint measurement to be as large as possible leads to the marginal entropy being a small as possible, for a given state.  

One can also try to find a similar result for the joint entropy, i.e. try to show that for a given state the joint entropy is minimised by performing an optimal joint measurement.  The difficulty with this is that the joint unbiasedness condition only specifies the form of the marginal probability distributions.  In principle many different joint probability distributions could lead to the same marginal distributions.  This means that it is difficult to find a general proof that optimal joint measurements minimise the joint entropy. It is possible, however, to prove that optimal joint measurements are best for a restricted class of joint measurements \cite{thesis}.

\section{A lower bound on the joint entropy}
\label{jents}

Our first task is to obtain a lower bound on the joint entropy, $H(A_J,B_J)$.  Let $\alpha$ and $\beta$ be the sharpnesses for a joint measurement of $\hat A={\bf a}\cdot\boldsymbol{ \hat \sigma}$ and $\hat B={\bf b}\cdot\boldsymbol{ \hat \sigma}$ respectively.  We use the measurement scheme outlined in \cite{busch2, erikas} and in section \ref{jmeasurements}.  This provides an optimal joint measurement in which the joint
probability distribution $P^{ab}_{ij}$ is 
\begin{eqnarray}
P^{ab}_{\pm,\pm}&=&pP^m_{\pm},\nonumber\\
P^{ab}_{+,-}&=&(1-p)P^l_+,\nonumber\\
P^{ab}_{-,+}&=&(1-p)P^l_-,
\end{eqnarray}
where $p$ is the probability of measuring along the ${\bf m}$ axis and where $P^m_{\pm}$ and $P^l_{\pm}$ are the probabilities of obtaining the result $\pm$ for a measurement of spin along ${\bf m}$
and ${\bf l}$ respectively.  It can be shown that the joint entropy will be
\begin{eqnarray}
\label{jentropy}
H(A_J,B_J)&=&-p\log(p)-(1-p)\log(1-p)\nonumber\\
&+&pH(M)+(1-p)H(L),
\end{eqnarray}
where $H(M)$ and $H(L)$ are the entropies for measurements of the spin along ${\bf m}$ and ${\bf l}$ respectively.  From equation (\ref{p}), we see that $p$ is specified completely by the choice of $\alpha{\bf a}$ and $\beta{\bf b}$.  This means that the only part of (\ref{jentropy}) that depends on the initial state is the convex sum 
\begin{equation}
\label{convex}
pH(M)+(1-p)H(L).
\end{equation}
We shall thus focus our attention on finding a state independent lower bound on this quantity. 

\subsection{Lower bound for the case when both observables are measured equally sharply}
If we wished to measure each spin component equally sharply then we would take $\alpha=\beta$.  For this situation a tight lower bound\footnote{By tight we mean that the lower bound can be saturated.} can be obtained for (\ref{convex}).  From equations (\ref{m}) and (\ref{l}) it is clear that ${\bf m\cdot l}=0$ when $\alpha=\beta$.  The observables ${\bf m\cdot\hat\sigma}$ and ${\bf l\cdot\hat\sigma}$ are thus complementary.  The Maassen and Uffink relation, (\ref{mu}), then becomes 
\begin{equation}
H(M)+H(L)\ge 1.
\end{equation}
This lower bound will be saturated when the state is prepared as an eigenstate of either ${\bf m}\cdot\hat\sigma$ or ${\bf l}\cdot\hat\sigma$.  If $p=(1-p)=1/2$, then it is clear
that the $pH(M)+(1-p)H(L)\ge p$, where this lower bound can be saturated.  Consider now the case when $p<1/2$ and consequently $(1-p)>p$.  We can re-express the convex sum of entropies as 
\begin{eqnarray}
\label{convex1}
pH(M)+(1-p)H(L)&=&p[H(M)+H(L)]\nonumber\\
+(1-2p)H(L)&\ge& p+(1-2p)H(L).
\end{eqnarray}
To obtain $H(M)+H(L)=1$, our state must be an eigenstate of either ${\bf m\cdot\hat\sigma}$ or ${\bf l\cdot\hat\sigma}$.  If we take the state to be an eigenstate of ${\bf l\cdot\hat\sigma}$, then $H(L)=0$.  We thus find that the minimum of $pH(M)+(1-p)H(L)$ is $p$, when $\alpha=\beta$ and $p<1/2$.  For the case when $p>1/2$, we can re-express the sum of the entropies as 
\begin{eqnarray}
\label{convex2}
pH(M)+(1-p)H(L)&=&(1-p)[H(M)+H(L)]\nonumber\\
&-&(1-2p)H(M).
\end{eqnarray}
It can be seen that the minimum of the convex sum of the entropies is $1-p$, which is obtained when the state is an eigenstate of ${\bf m\cdot\hat\sigma}$.  Collecting these
results we find that the lower bound on the joint entropy, for $\alpha=\beta$, is 
\begin{equation}
\label{exactlb}
H(A_J,B_J)\ge
\begin{cases}
H_2(p)+p & \text{if }p\le \frac{1}{2},\cr
H_2(p)+(1-p) &\text{if }p\ge \frac{1}{2},
\end{cases}
\end{equation}
where $H_2(x)=-x\log(x)-(1-x)\log(1-x)$.  When $p\le 1/2$, then the equality in this lower bound is obtained when the state is an eigenstate of ${\bf l\cdot\hat\sigma}$.  When $p\ge 1/2$, then the equality in the lower
bound is obtained when the state is an eigenstate of ${\bf m\cdot\hat\sigma}$.  The lower bound is expressed in terms of $p$, which refers to the specific measurement scheme
that is described in \cite{busch2, erikas} and in section \ref{jmeasurements}.  In equation (\ref{p}), both $p$ and $1-p$ are expressed in terms of $\alpha$, ${\bf a}$ and ${\bf
b}$.  It is thus possible to express the lower bound (\ref{exactlb}) in terms of $\alpha$, ${\bf a}$ and ${\bf b}$ as
\begin{equation}
\label{exactlb2}
H(A_J, B_J)\ge
\begin{cases}
H_2(\frac{\alpha}{2}|{\bf a+b}|)+\frac{\alpha}{2}|{\bf a-b}| & \text{   if   }\eta\le \frac{\pi}{2},\cr
H_2(\frac{\alpha}{2}|{\bf a+b}|)+\frac{\alpha}{2}|{\bf a+b}| & \text{   if   }\eta\ge \frac{\pi}{2},
\end{cases}
\end{equation}
where $\eta$ is the angle between ${\bf a}$ and ${\bf b}$.  It should be noted that the lower bound is only a function of ${\bf a}$
and ${\bf b}$.  The reason for this is that we are considering optimal joint measurements, which means that the value of $\alpha$ is fixed by (\ref{ineq}).  The lower bound thus depends only on the choice of spin components that we wish to jointly measure and not on the state.  

The state that achieves the minimum of $H(A_J,B_J)$, and
thus saturates the inequality (\ref{exactlb}) is an eigenstate of ${\bf m\cdot\hat\sigma}$ for $\eta\le\pi/2$ or an eigenstate of ${\bf l\cdot\hat\sigma}$ for $\eta\ge\pi/2$. 
Using equations (\ref{m}), (\ref{l}) and (\ref{p}) it can be shown that when $\alpha=\beta$
\begin{equation}
{\bf m}=\frac{{\bf a+b}}{|{\bf a+b}|},\;{\bf l}=\frac{{\bf a-b}}{|{\bf a-b}|}.
\end{equation}
Thus we have the following results.  When $\eta\le\pi/2$, the state that minimises the joint entropy will be a pure state with a Bloch vector that is confined to the plane
defined by ${\bf a}$ and ${\bf b}$, where the Bloch vector will lie exactly midway between ${\bf a}$ and ${\bf b}$.  For $\eta\ge\pi/2$ we have that the minimum entropy state
will again be a pure state with a Bloch vector that is confined to the ${\bf ab}$ plane, however, now the Bloch vector will be midway between ${\bf a}$ and $-{\bf b}$.  When the system is prepared in the minimum entropy state, we find that $H(A_J)=H(B_J)$.  It is interesting that the marginal entropies are equal when we minimise the joint entropy, $H(A_J,B_J)$.  From a physical perspective this observation seems to be a consequence of the fact that we want to measure each qubit observable equally well, i.e. $\alpha=\beta$.  We will now consider the case of joint measurements with $\alpha\ne\beta$. 

\subsection{Lower bound for the case when one observable is measured sharper that the other}

When $\alpha\ne\beta$, then ${\bf m\cdot\bf l}\ne 0$ and the spin observables ${\bf m\cdot\hat\sigma}$ and ${\bf l\cdot\hat\sigma}$ are not complementary.  To obtain the lower
bound we will again separate the analysis into two cases, when $p\le 1/2$ and when $p\ge1/2$.  For the first case, $p\le 1/2$, we can re-express the  convex sum of entropies as
$pH(M)+(1-p)H(L)=p[H(M)+H(L)]+(1-2p)H(L)$.  From (\ref{mu}), it is clear that  
\begin{eqnarray}
pH(M)+(1-p)H(L)\ge &-&2p\log|\langle m|l\rangle|_{max}\nonumber\\
&+&(1-2p)H(L), 
\end{eqnarray}
where $|\langle m|l\rangle|_{max}$ is the maximum overlap between the eigenvectors of ${\bf m\cdot\hat\sigma}$ and ${\bf l\cdot\hat\sigma}$.  The term $(1-2p)H(L)$ is 
non-negative for $p\le1/2$.  It follows that $-2p\log|\langle m|l\rangle|$ is a lower bound on $pH(M)+(1-p)H(L)$, when $p\le 1/2$.  For the other situation, where $p\ge 1/2$,
 we can re-express the sum of the entropies as $(1-p)[H(M)+H(L)]-(1-2p)H(M)$.  Using (\ref{mu}) and the fact that $(1-2p)H(M)$ is non-positive for $p\ge 1/2$, we
 obtain the following lower bound, $pH(M)+(1-p)H(L)\ge-2(1-p)\log|\langle m|l\rangle|_{max}$.  The lower bound on the joint entropy will thus be 
\begin{equation}
\label{lb1}
H(A_J, B_J)\ge 
\begin{cases}
H_2(p)-2p\log|\langle m|l\rangle|_{max} & \text{ if }p\le 1/2,\cr
H_2(p)-2(1-p)\log|\langle m|l\rangle|_{max} & \text{ if }p\ge 1/2.
\end{cases}
\end{equation}
This lower bound is expressed in terms of $p$ and $|\langle m|l\rangle|$, which both refer to one particular implementation of a joint measurement.  It would be better to
express the lower bound in terms of $\alpha$, $\beta$, ${\bf a}$ and ${\bf b}$, which apply to any joint measurement of two spin components.  In order for us to achieve this, we must note
that $|\langle m|l\rangle|^2_{max}=1/2(1+|{\bf m\cdot l}|)$.  By using equations (\ref{ineq}), (\ref{m}) and (\ref{l}) it can be shown that  
\begin{equation}
|\langle m|l\rangle|_{max}=\frac{1}{\sqrt{2}}\sqrt{1+\frac{|\alpha^2-\beta^2|}{2-\alpha^2-\beta^2}}.
\end{equation}
The lower bound (\ref{lb1}) may thus be re-written as
\begin{widetext}
\begin{equation}
\label{jmlb1}
H(A_J, B_J)\ge
\begin{cases}
H_2(\frac{1}{2}|\alpha{\bf a}+\beta{\bf b}|)-|\alpha{\bf a}-\beta{\bf b}|\log\left(\frac{1}{\sqrt{2}}\sqrt{1+\frac{|\alpha^2-\beta^2|}{2-\alpha^2-\beta^2}}\right) &\text{   if   }\eta\le\frac{\pi}{2},\cr
H_2(\frac{1}{2}|\alpha{\bf a}+\beta{\bf b}|)-|\alpha{\bf a}+\beta{\bf b}|\log\left(\frac{1}{\sqrt{2}}\sqrt{1+\frac{|\alpha^2-\beta^2|}{2-\alpha^2-\beta^2}}\right) &\text{   if   }\eta\ge\frac{\pi}{2}.
\end{cases}
\end{equation}
\end{widetext}
While we have derived equation (\ref{jmlb1}) for the specific example of the measurement scheme outlined in \cite{busch2,erikas}, the result will apply to any optimal joint measurement, i.e. one that satisfies the unbiasedness condition and saturates the inequality (\ref{ineq}).  When $\alpha\ne\beta$ the bound (\ref{jmlb1}) cannot be saturated.  One reason for this is that when we derived (\ref{jmlb1}) we made use of the relation (\ref{mu}).  It was
shown that for the case of two spin 1/2 particles, that equation (\ref{mu}) can only be saturated when the two spin components are orthogonal to each
other \cite{GMR}.  

\begin{figure}
\center{\includegraphics[width=8cm,height=!]{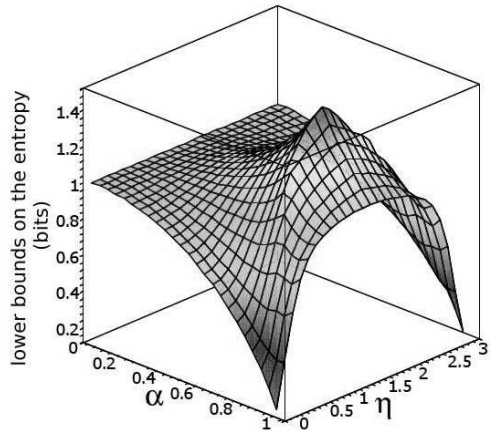}}
\caption{A plot of (\ref{jmlb1}) which shows the lower bound of $H(A_J, B_J)$, against $\alpha$ and $\eta$.  The value of $\beta$ is chosen to be as large as possible given the
choice of $\alpha$ and $\eta$.}
\label{entplot2}
\end{figure}

A plot of (\ref{jmlb1}) is given in figure \ref{entplot2}, where the lower bound is plotted against $\eta$
and $\alpha$.  The value of $\beta$ is chosen to be as large as equation (\ref{ineq}) allows, given the choice of $\alpha$ and $\eta$.  An interesting feature of figure \ref{entplot2} is that for any fixed choice of $\alpha$, the maximum value for $H(A_J,B_J)$ is obtained when $\eta=\pi/2$.  This result is physically reasonable because $\eta=\pi/2$ corresponds to the two spin components being complementary to each other.  The entropic uncertainty principle thus shows that we are most uncertain of the outcome when the two spin components are complementary.  %Similarly, figure \ref{entplot2} shows that for a fixed value of $\alpha$, the minimum of $H(A_J,B_J)$ occurs when $\eta=0$ or $\pi$.  This corresponds to the two spin components being parallel or anti-parallel, i.e. ${\bf a}=\pm{\bf b}$.  
The entropic uncertainty principle, given in equation (\ref{jmlb1}), allows us to quantify the notion of complementarity for the case when we jointly measure the two qubit observables.

\section{Lower bounds on the sum of the marginal entropies}
\label{margents}

Entropic uncertainty principles are often formulated in terms of the sum of the marginal entropies \cite{deutsch, maassen}.  For this reason it is useful to find a lower bound on the sum of the marginal entropies for the case of joint measurements.  The lower bound in equation (\ref{jmlb1}) will act as a lower bound on both the joint entropy and the sum of the marginal entropies.  In order for this
lower bound to be tight with respect to $H(A_J)+H(B_J)$ we would require that $H(A_J)+H(B_J)=H(A_J,B_J)$, which will not be true in general.  If we wish to find
better bounds on the sum of the marginal entropies $H(A_J)+H(B_J)$, then we should work directly with this sum.  In this section we shall present two different approaches to bounding the sum of the entropies.  The first approach will be to modify an argument presented
in \cite{GMR} for obtaining the minimum of $H(A)+H(B)$ for two spin observables that are measured on separate but identically prepared systems.  The second approach will be to use the concavity of entropy to obtain a simple lower bound.

It is known that the bound of Maassen and Uffink, (\ref{mu}), can be saturated for complementary observables.  This raises the question of whether it can also be saturated for
observables that are  not complementary.  Numerical investigations for systems described in a two dimensional Hilbert space indicate that this is not the case \cite{GMR}. 
This has led to investigations into whether it is possible to improve upon equation (\ref{mu}) for systems in a two dimensional Hilbert space \cite{sanchezruiz, GMR}.  A simple
analytical lower bound, which could be saturated, was found in \cite{GMR}.  For the case of measurements of two spin components, with an angle of $\eta$ between them, the lower
bound of \cite{GMR} took the form
\begin{equation}
\label{GR}
H(A)+H(B)\ge
\begin{cases}
2H_2\left(\frac{1}{2}+\frac{1}{2}\cos(\eta/2)\right) &\text{ if }\eta\in [0, \eta'],\cr
2H_2\left(\frac{1}{2}+\frac{1}{2}\sin(\eta/2)\right) &\text{ if }\eta\in [\pi-\eta', \pi],
\end{cases}
\end{equation}
where $\eta'\approx 1.17056$ radians, and where $H_2(x)=-x\log(x)-(1-x)\log(1-x)$.  It can be seen that this lower bound only applies for a limited rangle of angles between the
two spin components.  Outwith this range the minimum was found numerically.  It is important to note that (\ref{GR}) was derived for the case of measurements of $\hat A$ and $\hat B$ on separate but identically prepared systems.  A similar result will now be derived for the case of joint measurements that are equally sharp, i.e. $\alpha=\beta$. 

If we have a qubit system prepared in the state $\hat\rho=1/2(\hat 1+{\bf c\cdot\hat\sigma})$, then the marginal probabilities for obtaining the results spin up or down for
$\hat A$ and $\hat B$ measured jointly are given by \cite{erikas}
\begin{eqnarray}
\label{marginalprob}
P^{\alpha a}_{\pm}=\frac{1}{2}\pm\frac{\alpha}{2}{\bf a\cdot c},\nonumber\\
P^{\beta b}_{\pm}=\frac{1}{2}\pm\frac{\beta}{2}{\bf b\cdot c}.
\end{eqnarray}
The entropic uncertainty $H(A_J)+H(B_J)$ will thus equal $H_2(1/2+\alpha/2{\bf a\cdot
c})+H_2(1/2+\beta/2{\bf b\cdot c})$, where $H_2(x)=-x\log(x)-(1-x)\log(1-x)$.  In order to minimise the entropy $H_2(x)$, we must make the terms $x$ and $(1-x)$ as different
from each other as possible.  It is clear that if $H(A_J)$ and $H(B_J)$ are to be minimised then we must maximise $|P^{\alpha a}_+-P^{\alpha a}_-|=|\alpha{\bf a\cdot c}|$ and
$|P^{\beta b}_+-P^{\beta b}_-|=|\beta{\bf b\cdot c}|$.  This suggests that we should take ${\bf c}$ to be confined to the plane of the Bloch sphere that is defined by the
vectors ${\bf a}$ and ${\bf b}$.  Furthermore we should take the magnitude of ${\bf c}$ to be as large as possible, which implies that $|{\bf c}|=1$.  To summarise, we know that
the state that minimises $H(A_J)+H(B_J)$ must be a pure state, with a Bloch vector that is confined to the plane defined by the vectors ${\bf a}$ and ${\bf b}$.  These observations greatly
simplify the problem as we have reduced the number of parameters that we need to minimise over from three to just one.  We will parameterise the state
$\hat\rho=|\psi\rangle\langle\psi|$ as $|\psi\rangle=\cos(\theta/2)|a+\rangle+\sin(\theta/2)|a-\rangle$, where ${\bf a\cdot\hat\sigma}|a\pm\rangle=\pm|a\pm\rangle$.  The
variable $\theta$ is the angle between ${\bf c}$ and ${\bf a}$.  The marginal probabilities will now be $P^{\alpha a}_{\pm}=1/2\pm\alpha/2\cos(\theta)$ and $P^{\beta
b}_{\pm}=1/2\pm\beta/2\cos(\eta-\theta)$.  We shall now look at the situation where we measure $\hat A$ and $\hat B$ equally well and thus $\alpha=\beta$.  

\subsection{Lower bound for the case when both observables are measured equally sharply}

When $\alpha=\beta$, then the largest value for $\alpha$ allowed by equation (\ref{ineq}) is $\sqrt{1/(1+|\sin \eta|)}$.  Numerical
investigations show that $H(A_J)+H(B_J)$ assumes a global minimum at $\theta=\eta/2$, for a large range of values for $\eta$.  When $\eta$ approaches $\pi/2$ then the minimum
changes to a maximum.  It is straightforward to show that 
\begin{equation}
\frac{d}{d\theta}\left[H(A_J)+H(B_J)\right]_{\theta=\eta/2}=0. 
\end{equation}
Evaluating the second derivative of $H(A_J)+H(B_J)$ at $\theta=\eta/2$ yields 
\begin{eqnarray}
\label{secondder}
&\;&\frac{d^2}{d\theta^2}\left[H(A_J)+H(B_J)\right]_{\theta=\eta/2}=\nonumber\\
&\;&\frac{-\alpha}{\ln(2)}\left[\cos(\eta/2)\ln\left(\frac{1-\alpha\cos(\eta/2)}{1+\alpha\cos(\eta/2)}\right)+\frac{2\alpha\sin^2(\eta/2)}{1-\alpha^2\cos^2(\eta/2)}\right].
\end{eqnarray}
It can be seen that the second derivative, (\ref{secondder}), is positive at $\theta=\eta/2$ for $0\le\eta\le\eta'$.  The value of $\eta'$ is found by setting equation
(\ref{secondder}) equal to zero and solving for $\eta$.  By solving the equation numerically we find that $\eta'\approx 1.46117$.  We have thus found that when the angle
between ${\bf a}$ and ${\bf b}$ is in the range $0\le\eta\le\eta'$, where $\eta'\approx 1.46117$, then 
\begin{equation}
\label{exactjm1}
H(A_J)+H(B_J)\ge 2H_2\left(\frac{1}{2}+\frac{\alpha}{2}\cos(\eta/2)\right).
\end{equation}
This lower bound is optimal, in the sense that it can be saturated.  

\begin{figure}
\center{\includegraphics[width=11cm, height=!]{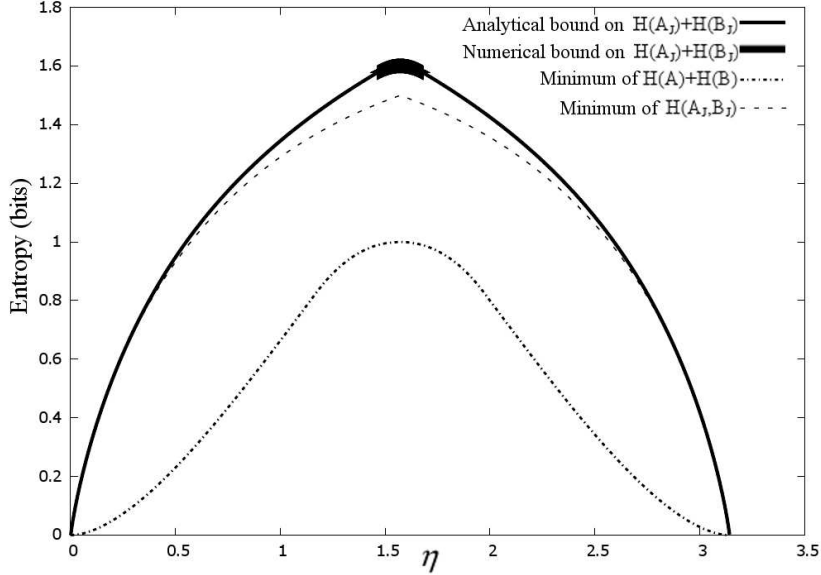}}
\caption{A plot showing the minimum of $H(A_J)+H(B_J)$ plotted against $\eta$, for $\alpha=\beta$.  The thick black line represents numerical results, while the thin black line is a plot of the analytical
results equations (\ref{exactjm1}) and (\ref{exactjm2}).  The dashed line is the minimum of the joint entropy $H(A_J, B_J)$, which for $\alpha=\beta$ is given by equation (\ref{exactlb2}).  The dotted line is
the minimum of $H(A)+H(B)$, evaluated numerically, which is the entropy for measurements of $\hat A$ and $\hat B$ made on separate copies of the same state.}
\label{minjent}
\end{figure}

For $\eta$ in the range $\pi\ge\eta\ge\pi-\eta'$ it can be shown that the entropy is minimal
when the vector ${\bf c}$ is midway between the vectors ${\bf b}$ and $-{\bf a}$, and thus $\theta=\eta/2+\pi/2$.  This last fact may be established using an argument that is similar
to the one that we employed to show that $\theta=\eta/2$ was a minimum for $0\le\eta\le\eta'$.  This leads to the result that when $\eta\in[\pi-\eta', \pi]$, then 
\begin{equation}
\label{exactjm2}
H(A_J)+H(B_J)\ge 2H_2\left(\frac{1}{2}+\frac{\alpha}{2}\sin(\eta/2)\right).
\end{equation}
Again this lower bound is optimal in the sense that it is possible to saturate the inequality.  When $\eta$ is in the region $\eta'<\eta<\pi-\eta'$ the minimum entropy state splits in two minimum entropy states.  The position of these states also jumps discontinuously from midway between $\bf a$ and $\bf b$ (or $\bf a$ and $-\bf b$ when $\eta>\pi-\eta'$) to one of them being closer to $\bf a$ while the other is closer to $\bf b$.  While we do not have an analytical expression for the position of the minimum entropy state when $\eta\in(\eta', \pi-\eta')$, we can still find the
minimum entropy numerically.  One situation that is simple to solve analytically is when $\eta=\pi/2$.  It can be shown that the minimum occurs when $|\psi\rangle$ is an
eigenstate of either $\hat A$ or $\hat B$.  This leads to the result that $H(A_J)+H(B_J)\ge 1+H_2(1/2+\alpha/2)$.  For $\eta=\pi/2$, we have that $\alpha=1/\sqrt{2}$. 
The lower bound will thus be $H(A_J)+H(B_J)\ge 1+H_2(1/2+\alpha/2)\approx 1.60088$. A plot of the minimum of $H(A_J)+H(B_J)$ is shown in figure \ref{minjent}.  The plot shows
that the minimum of $H(A_J)+H(B_J)$ is always greater than the minimum of $H(A)+H(B)$, for separate measurements of $\hat A$ and $\hat B$.  Thus we see that the price
for performing a joint measurement is an increase in the uncertainty in the outcome.  Figure \ref{minjent} also shows that the maximum of $H(A_J)+H(B_J)$ occurs when
$\eta=\pi/2$, i.e. when the observables are complementary.  We can also see from figure \ref{minjent} that both $H(A_J)+H(B_J)$ and $H(A_J,B_J)$ are both zero only when $\eta=0$ and $\pi$.  This corresponds to the two spin components being parallel and anti-parallel, respectively.  It is clear that in this case $\hat A$ and $\hat B$ will commute. 

\subsection{Lower bound for the case when one observable is measured sharper that the other}
It is sometimes interesting to examine joint measurements where one observable is measured more sharply than the other.  For measurements such as this $\alpha\ne\beta$.  The approach
that we employed to derive equations (\ref{exactjm1}) and (\ref{exactjm2}) will not work in this instance due to the fact that the location of the minimum of $H(A_J)+H(B_J)$ is
not related in a simple manner to ${\bf a}$ and ${\bf b}$.  If we wish to place a lower bound on $H(A_J)+H(B_J)$ when $\alpha\ne\beta$, then from the subadditivity property of
entropy \cite{EIT, wehrl}, $H(A_J)+H(B_J)\ge H(A_J, B_J)$, and thus we may use the lower bound given by equation (\ref{jmlb1}).  Other entropic inequalities can, of course, be derived \cite{thesis}. However we shall only mention one more which follows as a simple consequence of the concavity of entropy.  It is clear that the marginal probabilities given in equation (\ref{marginalprob}) can be expressed as  
\begin{eqnarray}
\label{cgrained}
P^{\alpha a}_{\pm}&=&(1/2+\alpha/2)P^a_{\pm}+(1/2-\alpha/2)P^a_{\mp},\nonumber\\
P^{\beta b}_{\pm}&=&(1/2+\beta/2)P^b_{\pm}+(1/2-\beta/2)P^b_{\mp},
\end{eqnarray}
where $P^a_{\pm}$ and $P^b_{\pm}$ are the marginal probability distributions that one would obtain for sharp measurements of $\hat A$ and $\hat B$ respectively, on separate but identically prepared systems.  Using this simple fact together with the concavity of entropy \cite{EIT}, we find that 
\begin{eqnarray}
H_2(P^{\alpha a}_{\pm})&\ge& P^a_+H_2\left(\frac{1}{2}+\frac{\alpha}{2}\right)+P^a_-H_2\left(\frac{1}{2}-\frac{\alpha}{2}\right)\nonumber\\
&=&H_2\left(\frac{1}{2}+\frac{\alpha}{2}\right),\nonumber\\
H_2(P^{\beta b}_{\pm})&\ge& P^b_+H_2\left(\frac{1}{2}+\frac{\beta}{2}\right)+P^b_-H_2\left(\frac{1}{2}-\frac{\beta}{2}\right)\nonumber\\
&=&H_2\left(\frac{1}{2}+\frac{\beta}{2}\right),
\end{eqnarray}
and thus we obtain the inequality 
\begin{equation}
\label{jmlb4}
H(A_J)+H(B_J)\ge H_2\left(\frac{1}{2}+\frac{\alpha}{2}\right)+H_2\left(\frac{1}{2}+\frac{\beta}{2}\right).
\end{equation}
This last entropic uncertainty principle is worse than both (\ref{jmlb1}) and (\ref{exactjm1}) when $\alpha=\beta$.  For the case when we measure one observable better than the other, and thus $\alpha\ne\beta$, we find that equation (\ref{jmlb4}) is sometimes better than equation (\ref{jmlb1}) but not when $|\alpha-\beta|$ is small.  It is interesting to note that the lower bound in equation (\ref{jmlb4}) is equal to a quantity introduced by Martens and de Muynck to quantify the differences between joint measurements and separate measurements of two observables\footnote{To be more precise, Martens and de Muynck  were interested in joint measurements where the marginal POM elements for each of the jointly measured observables were a statistical mixture of the measurement operators for a sharp non-joint measurement of the observables.  In such cases one can think of the jointly measured observables as being non-ideal versions of the sharp observables.  The entropic quantity that they introduced was thus a measure of how non-ideal the jointly measured observables were.} \cite{martens, deMPOVM}.

\section{Comparison with an entropic uncertainty principle for POMs}

\begin{figure}
\center{\includegraphics[width=9cm,height=!]
{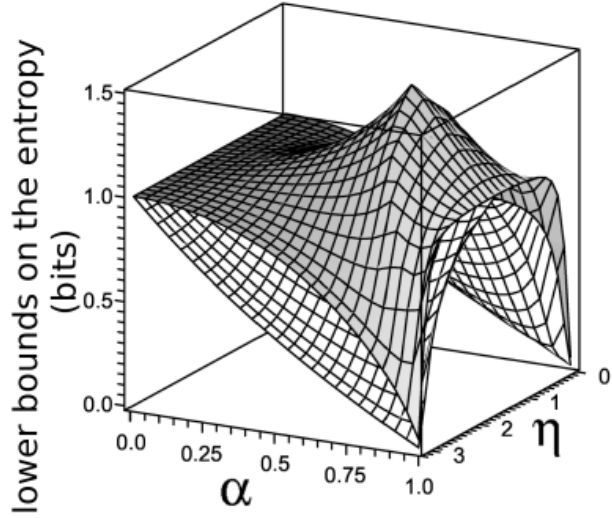}}
\caption{A plot comparing the lower bound (\ref{jmlb3}) to the lower bound on joint entropy given in equation (\ref{jmlb1}). The bounds are plotted against $\alpha$
and $\eta$, where $\beta$ has the largest value allowed from equation (\ref{ineq}).  The lower white surface represents the lower bound in equation (\ref{jmlb3}) while the darker surface is the lower bound given in equation (\ref{jmlb1}).}
\label{compare}
\end{figure}

The entropic uncertainty principle of Maassen and Uffink, (\ref{mu}), has been generalised to the case of POMs \cite{parth,mjh}.  In particular, Krishna and Parthasarathy obtained a generalised entropic uncertainty principle, which can be applied to POMs that can have rank greater than or equal to 1 \cite{parth}.  Their result states that for two POMs $\{\hat F_i\}$ and $\{\hat G_i\}$ 
\begin{equation}
\label{KP}
H(F)+H(G)\ge\max_{i,j}\left(-2\log|\hat F^{1/2}_i\hat G^{1/2}_j|\right),
\end{equation}
where $|\hat O|$ is the maximum of $\sqrt{\langle\psi|\hat O^{\dagger}\hat O|\psi\rangle}$, over all normalised $|\psi\rangle$.  It is worth comparing the results we have obtained with (\ref{KP}).  We must thus calculate the maximum of $|\sqrt{\hat\Pi^{\alpha a}_i}\sqrt{\hat\Pi^{\beta b}_j}|$, where $\{\hat\Pi^{\alpha a}_i\}$ and $\{\hat\Pi^{\beta b}_j\}$ are the POMs that give the marginal probabilities $P^{\alpha a}_i$ and $P^{\beta b}_j$ respectively.  It can be shown that 
\begin{eqnarray}
|\sqrt{\hat\Pi^{\alpha a}_+}\sqrt{\hat\Pi^{\beta b}_+}|&=&\frac{1}{2}\left[1+\alpha\beta\cos(\eta)+\sqrt{\alpha^2+\beta^2+2\alpha\beta\cos(\eta)-\alpha^2\beta^2\sin^2(\eta)}\right]^{1/2}\nonumber\\
|\sqrt{\hat\Pi^{\alpha a}_+}\sqrt{\hat\Pi^{\beta b}_-}|&=&\frac{1}{2}\left[1-\alpha\beta\cos(\eta)+\sqrt{\alpha^2+\beta^2-2\alpha\beta\cos(\eta)-\alpha^2\beta^2\sin^2(\eta)}\right]^{1/2}.
\end{eqnarray}
Similarly we can show by direct calculation that 
\begin{eqnarray}
|\sqrt{\hat\Pi^{\alpha a}_-}\sqrt{\hat\Pi^{\beta b}_+}|&=&|\sqrt{\hat\Pi^{\alpha a}_+}\sqrt{\hat\Pi^{\beta b}_-}|,\nonumber\\
|\sqrt{\hat\Pi^{\alpha a}_-}\sqrt{\hat\Pi^{\beta b}_-}|&=&|\sqrt{\hat\Pi^{\alpha a}_+}\sqrt{\hat\Pi^{\beta b}_+}|,\nonumber\\
|\sqrt{\hat\Pi^{\beta b}_i}\sqrt{\hat\Pi^{\alpha a}_j}|&=&|\sqrt{\hat\Pi^{\alpha a}_j}\sqrt{\hat\Pi^{\beta b}_i}|.
\end{eqnarray}
Using these results it is found that
\begin{widetext}
\begin{eqnarray}
\label{jmlb3}
H(A_J)+H(B_J)\ge\nonumber
\begin{cases}
-2\log\left\{\frac{1}{2} \left[1+\alpha\beta\cos(\eta)+\sqrt{\alpha^2+\beta^2+2\alpha\beta\cos(\eta)-\alpha^2\beta^2\sin^2(\eta)}\right]\right\} &\text{ if }\eta\le\frac{\pi}{2},\cr\cr
-2\log\left\{\frac{1}{2} \left[1-\alpha\beta\cos(\eta)+\sqrt{\alpha^2+\beta^2-2\alpha\beta\cos(\eta)-\alpha^2\beta^2\sin^2(\eta)}\right]\right\} &\text{ if }\eta\ge\frac{\pi}{2}.
\end{cases}\nonumber\\
\end{eqnarray}
\end{widetext}
%\begin{eqnarray}
%\label{jmlb3}
%H(A_J)+H(B_J)\ge\nonumber
%\end{eqnarray}
%\begin{eqnarray}
%\begin{cases}
%-2\log\left\{\frac{1}{2} \left[1+\alpha\beta\cos(\eta)+\sqrt{\alpha^2+\beta^2+2\alpha\beta\cos(\eta)-\alpha^2\beta^2\sin^2(\eta)}\right]\right\} &\text%{ if }\eta\le\frac{\pi}{2},\cr\cr
%-2\log\left\{\frac{1}{2} \left[1-\alpha\beta\cos(\eta)+\sqrt{\alpha^2+\beta^2-2\alpha\beta\cos(\eta)-\alpha^2\beta^2\sin^2(\eta)}\right]\right\} &\text%{ if }\eta\ge\frac{\pi}{2}.
%\end{cases}\nonumber\\
%\end{eqnarray}
A plot of the lower bound in equation (\ref{jmlb3}) is given in figure \ref{compare}, where it is plotted together with equation (\ref{jmlb1}).  It can be seen that the right hand side of the inequality (\ref{jmlb3}) is always less than our lower bound on $H(A,B)$, (\ref{jmlb1}).  We thus see that generalised uncertainty principle (\ref{KP}) does not provide as strong a bound on the uncertainty in a joint measurement as equation (\ref{jmlb1}).

\section{Conclusion} 
\label{conc}
We considered using entropy to quantify the uncertainty in the outcome of a joint measurement of two spin components of a spin 1/2 particle.  The reason for using entropy is that we can obtain nontrivial state independent lower bounds on the entropic uncertainty in the outcome of quantum measurements.  The fact that the lower bound does not depend on the state has important consequences.  For instance, state dependant lower bounds can sometimes be zero, even when the observables do not commute.  The state independent lower bounds that we obtained each depend only on the measurements that were being made.  It is this feature together with the fact that the lower bounds do not vanish for incompatible spin components, which make entropic uncertainty principles suitable as a mathematical formulation of the principle of complementarity.  

Two different entropic quantities were used to quantify the uncertainty.  These were the joint entropy, $H(A_J,B_J)$ and the sum of the marginal entropies, $H(A_J)+H(B_J)$.  One situation of particular interest was when we jointly measured both spin observables equally well, i.e. with equal sharpnesses.  For this case we were able to obtain lower bounds on the joint entropy, $H(A_J,B_J)$, which could be saturated.  %The state that achieved this was a pure state with a Bloch vector that was confined to the plane of the Bloch sphere defined by the unit vectors $\bf a$ and $\bf b$, which point along the directions of the two spin components that were jointly measured.  Specifically we found that when ${\bf a\cdot b}\ge 0$, the Bloch vector of the minimum entropy state was exactly midway between $\bf a$ and $\bf b$.  When ${\bf a\cdot b}\le 0$, the Bloch vector of the minimum entropy state was midway between $\bf a$ and $-\bf b$.  

We also obtained a lower bound on the sum of the marginal entropy, $H(A_J)+H(B_J)$, which could be saturated, for the case of measuring each observable equally well.  This lower bound, however, is only valid when the angle between the two spin components that were jointly measured lay in the rangle $[0, \eta']\cup[\pi-\eta', \pi]$, where $\eta'\approx 1.46117$ radians.  %When the angle between $\bf a$ and $\bf b$ was in this range, the minimum entropy state was a pure state with a Bloch vector that was confined to the plane defined by $\bf a$ and $\bf b$.  Specifically, when ${\bf a\cdot b}\ge 0$, the Bloch vector of the minimum entropy state was exactly midway between $\bf a$ and $\bf b$.  Alternatively, when ${\bf a\cdot b}\le 0$, the Bloch vector of the minimum entropy state was midway between $\bf a$ and $-\bf b$.  When the angle between $\bf a$ and $\bf b$ becomes greater than $\eta'$ but still less than $\pi-\eta'$, then we obtain two minimum entropy states, neither of which are exactly midway between $\bf a$ and $\bf b$.  

An important feature of the lower bounds on $H(A_J,B_J)$ and $H(A_J)+H(B_J)$, with $\alpha=\beta$, was that they could be saturated.  The states that achieves these lower bounds could then be obtained.  It was found that for $\eta\in[0,\eta']\cup[\pi-\eta']$, where $\eta'\approx 1.46117$ radians, the minimum entropy state was the same for both the joint entropy and the sum of the marginal entropies.  Furthermore, $H(A_J)=H(B_J)$ for the state that minimised the entropy.  When $\eta$ was greater than $\eta'$, but less than $\pi-\eta'$, the state that minimised $H(A_J)+H(B_J)$ was different from that which minimised the joint entropy.  Numerical investigations found that the minimum entropy state for $H(A_J)+H(B_J)$ split into two states, one of which had a Bloch vector that was close to ${\bf a}$, while the other had a Bloch vector that was close to ${\bf b}$.  This resulted in $H(A_J)\ne H(B_J)$, for the states that minimised the sum of the marginal entropies.  From a physical perspective this behaviour seems to be caused by the fact that as $\eta$ increases, the two spin components become more incompatible.  This causes the uncertainty to increase as $\eta$ increases.  Eventually we reach a point were we cannot minimise the sum of the marginal entropies by simultaneously having both $H(A_J)$ and $H(B_J)$ small.  Instead we must make one of the marginal entropies smaller than the other.  It is interesting that this behaviour does not occur for the joint entropy.  In fact the state that minimises $H(A_J, B_J)$ always results in $H(A_J)=H(B_J)$.

A lower bound (\ref{jmlb1}) was also obtained on the joint entropy for the case when we measured one observable more sharply than the other.  This lower bound, however, cannot be saturated.  The fact that entropy is subadditive means that the lower bound on the joint entropy will also act as a lower bound on the sum of the marginal entropies.  This lower bound was compared to that which was provided by the generalised entropic uncertainty principle of Krishna and Parthasarathy \cite{parth}.  It was found that our lower bound was always better than the lower bound obtained using the generalised entropic uncertainty principle.  

One approach that we have not discussed is using information Bell inequalities \cite{infobell} as a means of bounding the uncertainty of a joint measurement.  This approach is similar to that discussed in \cite{thesis} with regards to using information Bell inequalities to bound the sharpness of a joint measurement.  %This approach would be similar to that discussed in \cite{wonmin}.  Unfortunately, it seems that information Bell inequalities do not provide a tight bound on the sharpness of a joint measurement \cite{thesis}.  The difficulty lies in the fact that the size of the violations of the information Bell inequalities is often small and that the range of angles between spin components, over which they are violated, is less than for conventional Bell inequalities.   
Unfortunately, this approach does not lead to a tight bound on the uncertainty of a joint measurement.  The difficulty lies in the fact that the information Bell inequalities are only violated by entangled states.  In particular, maximally entangled states give the maximum violation.  If we have an entangled state, then the single particle reduced states will be mixed states.  The concavity of the entropy, however, means that the entropy is minimised only for pure states.

We have not addressed the problem of finding entropic uncertainty principles in higher dimensions.  One difficulty with this is that optimal joint measurements are not know for systems with dimension greater than 2.  It is still possible to construct joint measurements for obsevables in higher dimensions, and hence to find uncertainty principles for these measurements.  The difficulty with this, however, is that any lower bound on the uncertainty would apply only to one specific measurement.  If this measurement is not optimal, then one could in principle perform a different joint measurement that has less uncertainty than the given lower bound.

It is important to note is that the entropic uncertainty principles that we have obtained contain uncertainty from two different sources.  Firstly we have the inherent uncertainty in preparation that is quantified by the standard entropic uncertainty principles (\ref{mu}).  This uncertainty reflects the fact that we cannot prepare a state where we are certain of the outcome of separate measurements of both $\hat A$ or $\hat B$.  Secondly, there is a contribution to the uncertainty in our entropic uncertainty principles caused by the joint measurement.  One can thus view the difference between the lower bounds in our entropic uncertainty principles, and the minimum of $H(A)+H(B)$ for separate measurements of $\hat A$ and $\hat B$, as giving a coarse measure of the uncertainty that must be added in order to jointly measure the observables $\hat A$ and $\hat B$.

\acknowledgments
We would like to thank M. J. Hall for helpful discussions.  T.B. acknowledges financial support from the Doppler Institute grant LC 06002.  E.A. thanks the Royal Society for financial support.  S.M.B. thanks the Royal Society and the Wolfson Foundation for financial support.  We also acknowledge financial support from Czech grants MSMTL 06002 and MSM 6840770039  and from the Royal Society International Joint Project scheme.


\begin{thebibliography}{00}
\bibitem{rob} H. P. Robertson, Phys. Rev. {\bf 34}, 136 (1929). 
\bibitem{hirsch} I. I. Hirschman, Am. J. Math. {\bf 79}, 152 (1957).
\bibitem{bialynicki} I. Bialynicki-Birula and J. Mycielski, Commun. Math. Phys. {\bf 44}, 129 (1975).
\bibitem{deutsch} D. ~Deutsch, Phys.\ Rev.\ Lett. {\bf 50}, 631 (1983).
\bibitem{maassen} H. ~Maassen and J. ~B. ~M. ~Uffink, Phys.\ Rev.\ Lett. {\bf 60}, 1103 (1988).
\bibitem{massar} S. Massar, {Phys. Rev. A} {\bf 76}, 042114 (2007).
\bibitem{nch} M. A. Nielsen and I. L. Chuang, Quantum Computation and Quantum Information, Cambridge University Press, Cambridge, 2000.
\bibitem{peres} A. Peres, Quantum Theory; Concepts and Methods, Kluwer Academic Publishers, Dordrecht, 1998.
\bibitem{AK} E. Arthurs and J. L. Kelly, { Bell Syst. Tech. J.} {\bf 44}, 725 (1965).
\bibitem{AG} E. Arthurs and M. S. Goodman, { Phys. Rev. Lett. } {\bf 60}, 2447 (1988).
\bibitem{erikas} E. ~Andersson, S. ~M. ~Barnett and A. ~Aspect,  Phys. Rev. A {\bf 72}, 042104 (2005).
\bibitem{crypto} N. Gisin, G. Ribordy, W. Tittel, and H. Zbinden, {Rev. Mod. Phys.}, {\bf 74}, 145 (2002).
\bibitem{walker} N. G. Walker and J. E. Carrol, {Opt. Quantum Electron} {\bf 18}, 355 (1986).
\bibitem{stig} S. Stenholm, {Ann. Phys.} {\bf 218}, 233 (1992).
\bibitem{aperes} A. Peres, Phys. Rev. A 61, 022116 (2000).
\bibitem{busch}  P. ~Busch, Phys.\ Rev.\ D {\bf 33}, 2253 (1986); P. Busch, M. Grabowski, P. J. Lahti, Operational quantum physics, Springer-Verlag, Berlin, 1995.
\bibitem{steves} S. M. Barnett, { Phil. Trans. Roy. Soc. Lond. A} {\bf 59}, 1844 (1996).
\bibitem{busch2} P. ~Busch, Found.\ Phys. {\bf 17}, 905 (1987).
\bibitem{demuynck} W. M. de Muynck and H. Martens, Phys. Lett. A {\bf 142}, 187 (1989).
%\bibitem{opQM} P. Busch, M. Grabowski and P. J. Lahti, Operational quantum physics, Springer-Verlag, Berlin, 1995. 
\bibitem{hall} M. J. W. Hall, { Phys. Rev. A} {\bf 69}, 052113 (2004).
\bibitem{teiko} P. Busch, T. Heinonen, e-print, arXiv:0706.141 (2007).
\bibitem{coexist1} P. Busch, H. Schmidt, e-print, arXiv:0802.4167 (2008).
\bibitem{coexist2} P. Stano, D. Reitzner and T. Heinosaari, e-print, arXiv:0802.4248 (2008).
\bibitem{jmqubit} S. Yu, N. Liu, L. Li and C. H. Oh, e-print, arXiv:0805.1538 (2008).
\bibitem{cloning} T. Brougham, E. Andersson and S. M. Barnett,{ Phys. Rev. A} {\bf 73}, 062319 (2006).
\bibitem{uffink} J. Uffink, {In. J. Theor. Phys.} {\bf 33} 199 (1994).
\bibitem{estimation} T. Brougham and E. Andersson, {Phys. Rev. A} {\bf 76},  052313 (2007).
\bibitem{GMR} G. Ghirardi, L. Marinatto and R. Romano, { Phys. Lett. A}, {\bf 317}, 32 (2003).
\bibitem{sanchezruiz} J. Sanchez-Ruiz, {Phys. Lett. A}, {\bf 244}, 189 (1998).  
\bibitem{sruiz} J. Sanchez-Ruiz, {Phys. Lett. A}, {\bf 173}, 233, (1993).
\bibitem{winter} S. Wehner and A. Winter, {J. Math. Phys.} {\bf 49}, 062105 (2008).
\bibitem{EIT} T. ~M. ~Cover and J. ~A. ~Thomas, Elements of Information Theory, John Wiley and Sons, 1991.
\bibitem{wehrl} A. Wehrl, Rev.\ Mod.\ Phys. {\bf 50}, 221 (1978).
\bibitem{thesis} T. Brougham, Joint measurements on qubits and measurement correlation.  PhD thesis, Strathclyde University, Glasgow, U.K., 2008.  
\bibitem{martens} H. Martens and W. M. de Muynck, { Found. of Phys.}, {\bf 20}, 357 (1990).
\bibitem{deMPOVM} W. M. de Muynck, { Proceedings of the Workshop `Beyond the quantum', Leiden}, edited by T. M. Nieuwenhuizen, B. Mehmani, V. \v{S}picka, M. J. Aghdami and A. Y. Khrennikov, World Scientific, 2007; e-print quant-ph/0608087.
\bibitem{parth} M. Krishna and K. R. Parthasarathy, Phys. Indian Journal of Statistics A, {\bf 64}, 842 (2002).
\bibitem{mjh} M. J. W. Hall, {Phys. Rev A} {\bf 55}, 100 (1997).
\bibitem{infobell} S. L. Braunstein and C. M. Caves, {Phys. Rev. Lett.} {\bf 61}, 662 (1988).
%\bibitem{wonmin} W. Son, E. Andersson, S. M. Barnett and M. S. Kim, {\it Phys. Rev. A} {\bf 72}, 052116 (2005).
\end{thebibliography}
\end{document}